\documentclass[twocolumn,10pt]{IEEEtran}

\usepackage{cite}

\usepackage{graphicx}
\usepackage{amsmath}
\usepackage{amsfonts}
\usepackage{epstopdf}
\usepackage{xcolor}

\usepackage{algorithm}
\usepackage{algorithmic}
\usepackage{multirow}
\usepackage{transparent}
\usepackage{subfigure}
\usepackage{float}

\newtheorem{remark}{Remark}
\newtheorem{theorem}{Theorem}
\newtheorem{corollary}{Corollary}

\begin{document}
\title{Channel Characterization for 1D Molecular Communication with Two Absorbing Receivers}
\author{Xinyu Huang\thanks{X. Huang, N. Yang are with the Research School of Electrical, Energy and Materials Engineering, Australian National University, Canberra, ACT 2600, Australia (e-mail: \{xinyu.huang1, nan.yang\}@anu.edu.au).}, \textit{Student Member, IEEE}, Yuting Fang\thanks{Y. Fang is with the Department of Electrical and Electronic Engineering, University of Melbourne, Parkville, VIC 2010, Australia (e-mail: yuting.fang@unimelb.edu.au).}, Adam Noel, \textit{Member, IEEE}\thanks{A. Noel is with the School of Engineering, University of Warwick, Coventry, CV 7AL, UK (e-mail: adam.noel@warwick.ac.uk).}, \\and Nan Yang, \textit{Senior Member, IEEE}\vspace{-8mm}}
\maketitle
\begin{abstract}
	This letter develops a one-dimensional (1D) diffusion-based molecular communication system to analyze channel responses between a single transmitter (TX) and two fully-absorbing receivers (RXs). Incorporating molecular degradation in the environment, rigorous analytical formulas for i) the fraction of molecules absorbed, ii) the corresponding hitting rate, and iii) the asymptotic fraction of absorbed molecules as time approaches infinity at each RX are derived when an impulse of molecules are released at the TX. By using particle-based simulations, the derived analytical expressions are validated. Simulations also present the distance ranges of two RXs that do not impact molecular absorption of each other, and demonstrate that the mutual influence of two active RXs reduces with the increase in the degradation rate.
\end{abstract}
\vspace{-4mm}
\section{Introduction}

Molecular communication (MC) is one of the most promising solutions to nano-scale communications. In MC, information is encoded into small particles that are released by a transmitter ($\mathrm{TX}$) into a fluid medium and propagate until they arrive at a receiver $(\mathrm{RX})$. Moreover, MC can be biocompatible and consumes low energy.
These characteristics make MC suitable for applications such as targeted drug delivery, pollution control, and environmental monitoring \cite{nakano2012molecular}. For each application, accurate channel modeling is essential for analysis and design of MC systems \cite{jamali2019channel}.

Most existing MC papers have focused on the modeling of a single-$\mathrm{RX}$ MC system \cite{farsad2016comprehensive}. Some papers, e.g., \cite{fang2017convex,lu2016effect,guo2016eavesdropper,2019arXiv190307865K}, have considered a multi-$\mathrm{RX}$ MC system. The majority of papers involving a multi-$\mathrm{RX}$ MC system have assumed transparent $\mathrm{RXs}$ for tractability, due to the independence among  observations at multiple transparent $\mathrm{RXs}$. However, many practical $\mathrm{RX}$ surfaces might interact with the molecules of interest, e.g., by providing binding sites for absorption or other reactions \cite{cuatrecasas1974membrane}. In an environment where multiple non-transparent $\mathrm{RXs}$ co-exist, one non-transparent $\mathrm{RX}$ would impact molecules received by other non-transparent $\mathrm{RXs}$. Hence, an accurate characterization of such dependence makes the derivation of channel response (CR) cumbersome. Motivated by this, \cite{lu2016effect,guo2016eavesdropper,2019arXiv190307865K} have considered a multi-$\mathrm{RX}$ system with non-transparent $\mathrm{RXs}$. 
In \cite{lu2016effect}, the capture probability for each receiver was obtained via simulations. Considering a one-dimensional (1D) environment with two fully-absorbing $\mathrm{RXs}$, \cite{guo2016eavesdropper} derived the sum of absorbed molecules by both $\mathrm{RXs}$. Notably, \cite{2019arXiv190307865K} derived the fraction of molecules absorbed at each $\mathrm{RX}$ in a three-dimensional (3D) environment with two fully-absorbing $\mathrm{RXs}$. However, this derivation is not applicable in a 1D environment, as we will show in Section \ref{numerical}. Thus, an exact closed-form expression for the fraction of molecules absorbed over time at each $\mathrm{RX}$ (i.e., the CR) has not been derived yet for a 1D environment. 

Despite the aforementioned challenges, we provide closed-form expressions for the fraction of molecules absorbed at $\mathrm{RXs}$ when multiple fully-absorbing $\mathrm{RXs}$ co-exist, by taking into account the mutual influence between $\mathrm{RXs}$. Such expressions accurately characterize the CR at fully-absorbing $\mathrm{RXs}$ and lay the foundation for future performance evaluation, detection design, and diverse applications (e.g., target detection using two fully-absorbing $\mathrm{RXs}$) of a realistic multi-$\mathrm{RX}$ system. In this letter, we consider a 1D environment where one $\mathrm{TX}$ communicates between two fully-absorbing $\mathrm{RXs}$. The 1D environment is worthy of investigation since it is a good first approximation for regions between two close cells, such as chemical synapses in a human body \cite{keynes2001nerve}. To capture the effect of molecular chemical reaction on the received molecules at fully-absorbing $\mathrm{RXs}$, we also consider molecular degradation in the environment.

Our major contributions are summarized as follows. We derive i) the exact closed-form expressions for the fraction of molecules absorbed, ii) the corresponding hitting rate, and iii) the asymptotic fraction of molecules absorbed as time approaches infinity at each $\mathrm{RX}$ with an impulse emission at the $\mathrm{TX}$. Aided by a particle-based simulation method, we verify our analytical results. In addition, we present the distance ranges of two $\mathrm{RXs}$ that do not impact molecular absorption of each other. We also show that the mutual impact between two $\mathrm{RXs}$ reduces with the increase of degradation rate in the environment.

The rest of this paper is organized as follows. In Section II, we introduce the system model. In Section III, the closed-form expression for the fraction of absorbed molecules is derived. We also derive the corresponding hitting rate and the asymptotic fraction of absorbed molecules as time approaches infinity. In Section IV, we discuss the numerical results, and conclusion is presented in Section V.

\vspace{-4mm}
\section{System Model}\label{sec:system}
\begin{figure}[!t]
	\begin{center}
		\includegraphics[height=1in,width=1\columnwidth]{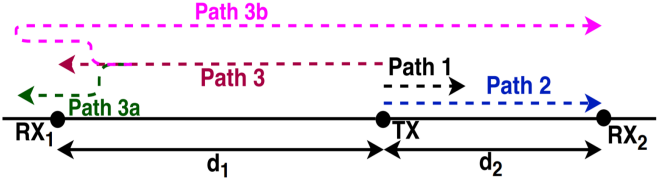}
		\caption{Illustration of the system model, where one TX communicates with two fully-absorbing RXs in a one-dimensional environment.}\label{system_model_figure}\vspace{-0.5em}
	\end{center}
\vspace{-4mm}
\end{figure}
In this letter, we consider a 1D unbounded environment where a single $\mathrm{TX}$ is located between two fully-absorbing $\mathrm{RXs}$, i.e., $\mathrm{RX_1}$ and $\mathrm{RX_2}$, with distance $d_1$ from the $\mathrm{RX_1}$ and distance $d_2$ from the $\mathrm{RX_2}$, as depicted in Fig. \ref{system_model_figure}. We consider the $\mathrm{TX}$ as a point source that can release an impulse of particles. We assume that the $\mathrm{TX}$ transmission starts at $t=0\;\mathrm{s}$.
Once released, particles diffuse randomly with a constant diffusion coefficient $D$. 
We also consider the first-order chemical reaction (i.e., unimolecular degradation) in the environment, where type $A$ molecules degrade into a new type of molecule $\phi$ that cannot be identified by the $\mathrm{RXs}$, i.e., $A\stackrel{k}{\longrightarrow}\phi$ \cite[Ch. 9]{chang2005physical}, where $k$ $[\mathrm{s^{-1}}]$ is the degradation rate constant. We model the two $\mathrm{RXs}$ as point fully-absorbing $\mathrm{RXs}$, which means that information molecules $A$ are absorbed as soon as they hit the point $\mathrm{RX_1}$ or $\mathrm{RX_2}$. 

\section{Derivation of Channel Impulse Response}\label{channel_response_section}
In this section, we derive closed-form expressions for the expected fraction of absorbed molecules at each $\mathrm{RX}$ for impulsive emission at the $\mathrm{TX}$, and the asymptotic fraction of absorbed molecules at each $\mathrm{RX}$ as $t\rightarrow\infty$. 
We first derive the CR for impulsive emission when one $\mathrm{RX}$ exists, which builds the foundation for deriving the CR when two $\mathrm{RXs}$ exist. According to \cite{heren2015effect}, the hitting rate for a single $\mathrm{RX}$ with molecular degradation can be obtained via the hitting rate without molecular degradation and multiplying by $\exp(-kt)$. Based on the existing expression for the hitting rate without molecular degradation in \cite{guo2016eavesdropper}, the hitting rate at the $\mathrm{RX}$ at time $t$, denoted by $f(d,t)$, with molecular degradation is
\begin{align}
f(d,t)=\frac{d}{\sqrt{4\pi Dt^3}}\exp\left(-\frac{d^2}{4Dt}-kt\right),
\end{align}
where $d$ is the distance between the $\mathrm{TX}$ and the $\mathrm{RX}$. Using $\int_{0}^{t}f(d,u)\mathrm{d}u$, we obtain the fraction of molecules absorbed by time $t$, denoted by $F(d,t)$, as
\begin{align}\label{fraction_single_degradation} F(d,t)=&\frac{1}{2}\exp\left(-\sqrt{\frac{k}{D}}d\right)\mathrm{erfc}\left(\frac{d}{\sqrt{4Dt}}-\sqrt{kt}\right)\notag\\&\!+\!\frac{1}{2}\exp\left(\sqrt{\frac{k}{D}}d\right)\mathrm{erfc}\left(\frac{d}{\sqrt{4Dt}}\!+\sqrt{kt}\right). 
\end{align}

As $t\rightarrow\infty$, we derive the asymptotic absorbed molecules $F(d,t\rightarrow\infty)$ as
\begin{align}\label{F_asy}
F(d,t\rightarrow\infty)=\exp\left(-\sqrt{\frac{k}{D}}d\right). 
\end{align}

In the following, we derive the CR when two $\mathrm{RXs}$ exist. We denote the fraction of absorbed molecules at $\mathrm{RX_1}$ and $\mathrm{RX_2}$ by time $t$ for impulsive emission by $P_{1}(t)$ and $P_{2}(t)$, respectively. We also denote the corresponding hitting rates at $\mathrm{RX_1}$ and $\mathrm{RX_2}$ by $p_{1}(t)$ and $p_{2}(t)$, respectively.

To derive $p_{1}(t)$ and $p_{2}(t)$, we first discuss the impact of the existence of $\mathrm{RX_1}$ on $p_{2}(t)$, based on \cite{2019arXiv190307865K}. As shown in Fig. \ref{system_model_figure}, we classify all possible diffusion paths of molecules by time $t$ in this environment into three paths, namely path 1, path 2, and path 3. Path 1 is for molecules diffusing in the environment, and path 2 and path 3 are for molecules that have hit $\mathrm{RX_2}$ and $\mathrm{RX_1}$, respectively. If only $\mathrm{RX}_2$ exists, we can further classify path 3 into path 3a and path 3b. Path 3a represents molecules that do not hit $\mathrm{RX}_2$ after firstly arriving at the location of $\mathrm{RX}_1$ at time $\tau<t$, and path 3b represents molecules that hit $\mathrm{RX}_2$ after firstly arriving at the location of $\mathrm{RX}_1$ at time $\tau$. Given that $f(d_2,t)$ denotes the hitting rate from the $\mathrm{TX}$ to $\mathrm{RX}_2$ when only $\mathrm{RX}_2$ exists, we find that $p_2(t)$ is less than $f(d_2,t)$, due to the existence of $\mathrm{RX}_1$. Accordingly, $p_2(t)$ is obtained as \cite[eq. (12)]{2019arXiv190307865K}
\begin{align}\label{n2}
p_2(t)=f(d_2,t)-\gamma(t),
\end{align}
where $\gamma(t)$ is the reduced hitting rate impacted by the existence of $\mathrm{RX}_1$. Based on the division in Fig. 1, $\gamma(t)$ is the hitting rate of path 3b and derived as
\begin{align}\label{gamma}
\gamma(t)=\int_{0}^{t}p_{1}(\tau)f(d_1+d_2,t-\tau)\mathrm{d}\tau,
\end{align}
where $f(d_1+d_2,t-\tau)$ is the hitting rate from $\mathrm{RX_1}$ to $\mathrm{RX_2}$ when $\mathrm{RX_1}$ is regarded as the $\mathrm{TX}$ and $\mathrm{RX_2}$ is the only $\mathrm{RX}$. Combining \eqref{n2} and \eqref{gamma}, we derive $p_{2}(t)$ as \cite[eq. (17)]{2019arXiv190307865K}
\begin{align}\label{hitting_rate_receiver_2}
p_{2}(t)=f(d_2,t)\!-\!\int_{0}^{t}\!{p_{1}(\tau)}f(d_1+d_2,t-\tau)\mathrm{d}\tau.
\end{align}

Similarly, we obtain $p_{1}(t)$ as
\begin{align}\label{hitting_rate_receiver_1}
p_{1}(t)=f(d_1,t)\!-\!\int_{0}^{t}\!p_{2}(\tau) f(d_1+d_2,t-\tau)\mathrm{d}\tau,
\end{align}
where $f(d_1,t)$ is the hitting rate from the $\mathrm{TX}$ to $\mathrm{RX_1}$ when only $\mathrm{RX_1}$ exists.

Based on \eqref{hitting_rate_receiver_2} and \eqref{hitting_rate_receiver_1}, we solve the closed-form expressions for $P_{2}(t)$ and $p_{2}(t)$ in the following theorem:
\begin{theorem}\label{theorem:n2}
The fraction of absorbed molecules at $\mathrm{RX_2}$ by time $t$ for an impulsive emission of molecules is given by
\begin{align}\label{absorbed_molecules_at_receiver_2}
P_{2}(t)=&\sum_{i=0}^{\infty}\left[R\left(2\left(i+1\right)d_1+\left(2i+3\right)d_2, t,  2\right)\right.\notag\\&\left.-R\left(2\left(i+2\right)d_1+\left(2i+3\right)d_2,t,2\right)\right.\notag\\&\left.-R\left(2id_1+\left(2i+1\right)d_2,t,0\right)\right.\notag\\&\left.+R\left(2(i+1)d_1+\left(2i+1\right)d_2,t,0\right)\right],
\end{align}
where $R(x,t,a)$ is given by
\begin{align}\label{P_1_deg}
R(x,t,a)=&\frac{\theta}{2}\sqrt{\frac{k}{D}}\left(\alpha\beta(t)-\tilde{\alpha}\tilde{\beta}(t)\right)
-\frac{i+1}{2}\Big(\tilde{\alpha}\tilde{\beta}(t)\Big.\notag\\&\Big.+\alpha\beta(t)\Big)-\frac{\theta}{\sqrt{\pi Dt}}\exp\left(-\frac{x^2}{4Dt}-kt\right)\notag\\&+(i+1).
\end{align}

In \eqref{P_1_deg}, $\theta=(d_1+d_2)(i+1)(i+a)$, $\alpha=\exp\left(x\sqrt{\frac{k}{D}}\right)$, $\tilde{\alpha}=\exp\left(-x\sqrt{\frac{k}{D}}\right)$, $\beta(t)=\mathrm{erfc}\left(\frac{x}{\sqrt{4Dt}}+\sqrt{kt}\right)$, and $\tilde{\beta}(t)=\mathrm{erfc}\left(\frac{x}{\sqrt{4Dt}}-\sqrt{kt}\right)$. The corresponding hitting rate at $\mathrm{RX_2}$ by time $t$, $p_{2}(t)$, is obtained by taking the derivative of \eqref{absorbed_molecules_at_receiver_2} with respect to $t$. By doing so, the expression for $p_2(t)$ is similar to \eqref{absorbed_molecules_at_receiver_2}, except for replacing $R(x,t,a)$ with $r(x,t,a)$, where $r(x,t,a)=\frac{\mathrm{d}R(x,t,a)}{\mathrm{d}t}$ and is given by
\begin{align}
r\left(x,t,a\right)=&\frac{i+1}{\sqrt{4\pi Dt^3}}\exp\left(-\frac{x^2}{4Dt}-kt\right)\notag\\&\times\left(\left(1-\frac{x^2}{2Dt}\right)(d_1+d_2)\left(i+a\right)-x\right).
\end{align}
\end{theorem}
\begin{IEEEproof}
Please see Appendix \ref{app}.
\end{IEEEproof}

We denote the asymptotic fraction of absorbed molecules as $t\rightarrow\infty$ at $\mathrm{RX}_1$ and $\mathrm{RX}_2$ by $P_{1,\mathrm{asy}}$ and $P_{2, \mathrm{asy}}$, respectively. We derive and present $P_{2, \mathrm{asy}}$ in the following theorem:

\begin{corollary}\label{theo2}
	The asymptotic fraction of absorbed molecules at $\mathrm{RX}_2$ as $t\rightarrow\infty$ is given by
	\begin{align}\label{asy}
	P_{2,\mathrm{asy}}&=\left\{\begin{array}{lr}
	\frac{\exp\left(-d_2\sqrt{\frac{k}{D}}\right)-\exp\left(-(2d_1+d_2)\sqrt{\frac{k}{D}}\right)}{1-\exp\left(-2(d_1+d_2)\sqrt{\frac{k}{D}}\right)}, k\neq 0\\
	\frac{d_1}{d_1+d_2}, k=0. 
	\end{array}
	\right.
	\end{align}
\end{corollary}
\begin{IEEEproof}
	Please see Appendix \ref{app2}.
\end{IEEEproof}

\begin{remark}
The closed-form expressions for $P_{1}(t)$, $p_{1}(t)$ and $P_{1, \mathrm{asy}}$ are obtained by exchanging $d_1$ and $d_2$ therein for $P_{2}(t)$, $p_{2}(t)$ and $P_{2, \mathrm{asy}}$, respectively.
\end{remark}
\begin{remark}
By setting $k=0$ in the expressions for $P_{2}(t)$ and $p_{2}(t)$, we can obtain the corresponding expressions without the occurrence of molecular degradation.
\end{remark}

\vspace{-2mm}
\section{Numerical Results}\label{numerical}

In this section, we present numerical results to validate our theoretical analysis in Section \ref{channel_response_section} and provide insightful discussions. The simulation results are conducted using a particle-based simulation method \cite{andrews2004stochastic}, where all results are averaged over 2000 realizations and the simulation time step is $\Delta t_{\mathrm{sim}}=0.001~\mathrm{s}.$ Throughout this section, we set the diffusion coefficient $D=79.4~\mu \mathrm{m}^2/\mathrm{s}$ \cite{yilmaz2014three}, an impulse of emission $N_{\mathrm{tx}}=5000$ molecules, $\Delta t=0.5~\mathrm{s}$, and $k=0.8~\mathrm{s^{-1}}$, unless otherwise stated. In all figures, we observe precise agreement between our simulation results and the analytical curves generated from Section \ref{channel_response_section}, which demonstrate the validity of our analysis. 

\begin{figure}[t]
	\centering
	\subfigure[Three data sets are applied: i) $d_1=d_2=20\;\mu\mathrm{m}$, $k=0$, ii) $d_1=25\;\mu\mathrm{m}$, $d_2=15\;\mu\mathrm{m}$, $k=0$, iii) $d_1=d_2=20\;\mu\mathrm{m}$, $k=0.8\;\mathrm{s}^{-1}$, where $D=79.4\;\mu\mathrm{m}^2/\mathrm{s}$ is applied for three data sets.]{
		\begin{minipage}[t]{1\linewidth}
			\centering
			\includegraphics[height=1.25in,width=3.5in]{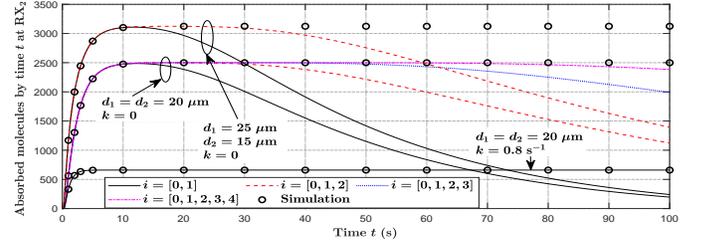}
			\label{a}
		\end{minipage}%
	}%
	\quad
	\subfigure[Two data sets are applied: i) $d_1=d_2=20\;\mu\mathrm{m}$, $D=79.4\;\mu\mathrm{m}^2/\mathrm{s}$, ii) $d_1=d_2=20\;\mu\mathrm{m}$, $D=100\;\mu\mathrm{m}^2/\mathrm{s}$, where $k=0$ is applied for two data sets.]{
		\begin{minipage}[t]{1\linewidth}
			\centering
			\includegraphics[height=1.25in,width=3.5in]{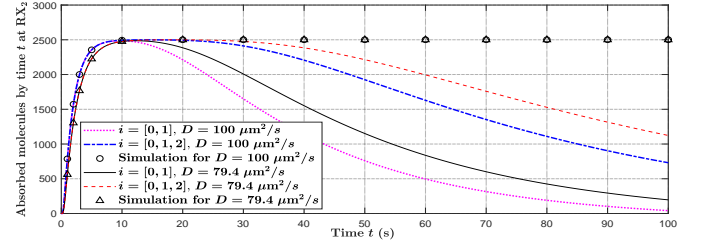}
			\label{c}
		\end{minipage}
	}
	
	\centering
	\caption{Absorbed molecules by time $t$ at $\mathrm{RX}_2$ versus time $t$, where different number of terms are applied in \eqref{absorbed_molecules_at_receiver_2}.}\label{term2}
\end{figure}

In Fig. \ref{term2}, we investigate the number of summation terms that should be applied in \eqref{absorbed_molecules_at_receiver_2}. In Fig. 2(a), we apply three data sets to investigate the impact of $d_1$, $d_2$, and $k$ on the number of summation terms. First, when only applying $i=\{0,1\}$ in \eqref{absorbed_molecules_at_receiver_2}, we observe that the equation first reaches the highest point, i.e., $P_{2,\mathrm{asy}}$, and then drops. Applying a larger number of terms results in $P_2(t)=P_{2,\mathrm{asy}}$ for a longer time, but increases computional complexity. To reduce such complexity, we clarify that applying $i=\{0,1\}$ is adequate since it enables \eqref{absorbed_molecules_at_receiver_2} to reveal the absorbed molecules before becoming visually indistinguishable from the asymptotic value, and increasing terms in \eqref{absorbed_molecules_at_receiver_2} does not change the absorbed molecules calculated before reaching the asymptotic value. After reaching the asymptotic value, we set $P_2(t)=P_\mathrm{asy}$. Second, comparing data set i) with data set ii) and data set iii), we observe that changing $d_1$, $d_2$, and $k$ does not change the fact that applying $i=\{0,1\}$ is adequate for \eqref{absorbed_molecules_at_receiver_2}. In Fig. 2(b), we apply two data sets to investigate the impact of $D$ on the number of terms. We still observe that changing $D$ does not impact the fact that applying $i=\{0,1\}$ is adequate for \eqref{absorbed_molecules_at_receiver_2}. 

\begin{figure}[!t]
	\begin{center}
		\includegraphics[height=2.5in,width=1\columnwidth]{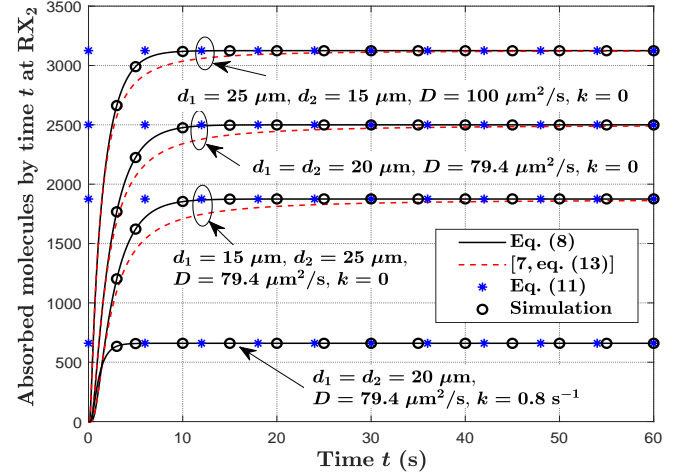}\vspace{-1em}
		\caption{Molecules absorbed by time $t$ at $\mathrm{RX_2}$ versus time $t$, where molecular emission is impulsive at the $\mathrm{TX}$.}\label{merge}\vspace{-1.5em}
	\end{center}
\end{figure}

In Fig. \ref{merge}, we plot molecules absorbed at $\mathrm{RX}_2$ by time $t$ using \eqref{absorbed_molecules_at_receiver_2} and \cite[eq. (13)]{2019arXiv190307865K}, respectively. We note that \cite[eq. (13)]{2019arXiv190307865K} was initially applied to a 3D environment but can also be applied to a 1D environment\footnote{In the 3D environment, \cite{2019arXiv190307865K} assumed that molecules are absorbed at the same points before reaching another $\mathrm{RX}$. The points on $\mathrm{RX}_1$ and $\mathrm{RX}_2$ are denoted by $s_1'$ and $s_2'$, where $s_1'$ and $s_2'$ are found numerically. As $\mathrm{RX}_1$ and $\mathrm{RX}_2$ are regarded as points in a 1D environment, $s_1'$ and $s_2'$ are points $\mathrm{RX}_1$ and $\mathrm{RX}_2$. Substituting $s_1'$ and $s_2'$ with $\mathrm{RX}_1$ and $\mathrm{RX}_2$ in \cite[eq. (13)]{2019arXiv190307865K}, we obtain the fraction of absorbed molecules at $\mathrm{RX}_2$ in a 1D environment, based on the method in \cite{2019arXiv190307865K}.}. In this figure, we first keep $k=0\;\mathrm{s}^{-1}$ and vary $d_1$, $d_2$, and $D$ to investigate the accuracy of \eqref{absorbed_molecules_at_receiver_2} and \cite[eq. (13)]{2019arXiv190307865K}. We also investigate the molecular degradation by setting $k=0.8\;\mathrm{s}^{-1}$ and plot the absorbed molecules at $\mathrm{RX}_2$ with \eqref{absorbed_molecules_at_receiver_2}. First, we clearly observe that the simulation matches well with \eqref{absorbed_molecules_at_receiver_2} and a gap exists between the simulation and \cite[eq. (13)]{2019arXiv190307865K} for $3\;\mathrm{s}\leq t\leq 30\;\mathrm{s}$, which demonstrates the accuracy advantage of (8) relative to \cite[eq. (13)]{2019arXiv190307865K}. Second, we observe that the asymptotic value of \eqref{absorbed_molecules_at_receiver_2} and \cite[eq. (13)]{2019arXiv190307865K} converge as $t\rightarrow\infty$. Last, we observe that \eqref{asy} matches with the simulation for both $k=0$ and $k=0.8\;\mathrm{s}^{-1}$ when $t$ is large, which demonstrates the correctness of \eqref{asy}.

\begin{figure}[!t]
	\begin{center}
		\includegraphics[height=2.5in,width=1\columnwidth]{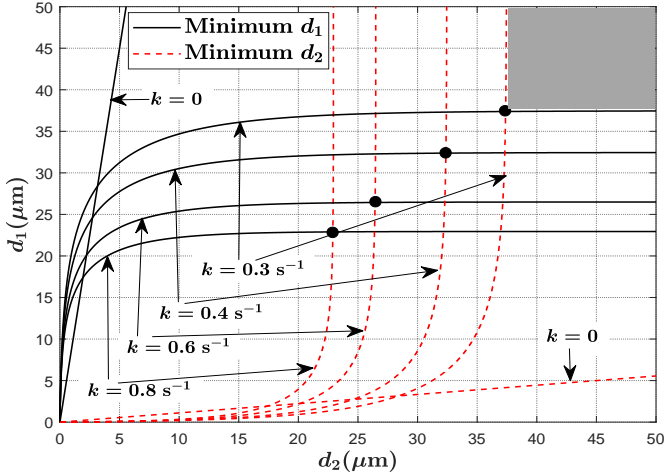}\vspace{-1em}
		\caption{Mimimum $d_1$ that does not influence absorbing molecules at $\mathrm{RX}_2$ versus $d_2$, and minimum $d_2$ that does not influence absorbing molecules at $\mathrm{RX}_1$ versus $d_1$, where different $k$ is applied. Constraint for calculating minimum $d_1$ and minimum $d_2$ is 0.01, and the grey area represents distances of two $\mathrm{RXs}$ that do not impact each other when $k=0.3\;\mathrm{s}^{-1}$.}\label{influence}
	\end{center}\vspace{-2em}
\end{figure}

In Fig. \ref{influence}, we plot the minimum $d_1$ that does not impact molecular absorption at $\mathrm{RX}_2$ versus $d_2$, and the minimum $d_2$ that does not impact molecular absorption at $\mathrm{RX}_1$ versus $d_1$, for different $k$. We examine the impact of $\mathrm{RX}_1$ on $\mathrm{RX}_2$ based on the gap between the fraction of absorbed molecules at $\mathrm{RX}_2$ and the fraction of absorbed molecules for the single $\mathrm{RX}$ as $t\rightarrow\infty$, which is expressed as $F(d_2,t\rightarrow\infty)-P_{2,\mathrm{asy}}$, where $F(d_2,t\rightarrow\infty)$ is given by \eqref{F_asy} and $P_{2,\mathrm{asy}}$ is given by \eqref{asy}. We calculate the minimum $d_1$ which satisfies the condition $\left(F(d_2,t\rightarrow\infty)-P_{2,\mathrm{asy}}\right)/F(d_2,t\rightarrow\infty)<0.01$ for given $d_2$. We also suppose that $\mathrm{RX}_1$ does not impact the molecular absorption at $\mathrm{RX}_2$ when this condition is satisfied. Similarly, we calculate the minimum $d_2$ for given $d_1$. From this figure, we first observe that minimum $d_1$ intersects minimum $d_2$ when $k\neq 0$. The upper right area of the intersection for each $k$ represents the range for $d_1$ and $d_2$ of two $\mathrm{RXs}$ that do not impact each other, because the condition for two $\mathrm{RXs}$ not impacting each other's molecular absorption is that $d_1$ and $d_2$ are simultaneously larger than minimum $d_1$ and minimum $d_2$, respectively. For example, if $d_1$ and $d_2$ are in the grey area, then two $\mathrm{RXs}$ do not impact each other when $k=0.3\;\mathrm{s}^{-1}$. When $k=0$, there is no intersection such that two $\mathrm{RXs}$ always impact each other. Second, we observe that the range for two $\mathrm{RXs}$ not impacting each other decreases with decreasing $k$. When $k$ decreases, there is a larger number of molecules in the environment such that the mutual influence between two $\mathrm{RXs}$ is higher, which results in the less range for two $\mathrm{RXs}$ not impacting each other. Third, we observe that minimum $d_1$ and minimum $d_2$ firstly increase when $d_2$ and $d_1$ increase, respectively, and then become constant after the intersection. This is because increasing $d_2$ means that the molecular absorption at $\mathrm{RX}_2$ decreases. In this case, if $d_1$ keeps the same value, then the molecular absorption at $\mathrm{RX}_1$ relatively increases, which results in a larger impact on the molecular absorption at $\mathrm{RX}_2$. Therefore, the minimum $d_1$ increases to reduce the impact on $\mathrm{RX}_2$. Beyond the intersection, two $\mathrm{RXs}$ will not impact each other such that increasing $d_2$ will not lead to increase in minimum $d_1$.

\vspace{-2mm}
\section{Conclusion}
In this letter, we focused on a 1D molecular communication system to investigate channel responses between a single $\mathrm{TX}$ and two fully-absorbing $\mathrm{RXs}$. We derived new closed-form expressions for i) the fraction of absorbed molecules, ii) the corresponding hitting rate, and iii) the asymptotic fraction of absorbed molecules as time approaches infinity at each $\mathrm{RX}$. Our results showed that our analytical expressions are accurate. We also investigated distance ranges for two $\mathrm{RXs}$ that do not impact molecular absorption of each other, which showed that the mutual influence between two $\mathrm{RXs}$ decreases with the increase in the degradation rate. Future work includes extending the 1D environment to 3D and deriving the CR between one $\mathrm{TX}$ and multiple $\mathrm{RXs}$ that partially absorb molecules.

\appendices

\section{Proof of Theorem \ref{theorem:n2}}\label{app}
	Taking the integral for both \eqref{hitting_rate_receiver_2} and \eqref{hitting_rate_receiver_1} over the interval $[0,t]$, we obtain

	\begin{align}\label{fraction_receiver_2}
		P_{2}(t)=F(d_2,t)-P_{1}(t)\ast f(d_1+d_2,t),
	\end{align}
	\begin{align}\label{fraction_receiver_1}
P_{1}(t)=F(d_1,t)-P_{2}(t)\ast f(d_1+d_2,t),
\end{align}	
where $*$ is the convolution operator. Substituting $P_1(t)$ in \eqref{fraction_receiver_2} with \eqref{fraction_receiver_1} and performing the Laplace transform, we obtain
\begin{align}\label{laplace_transform}
\mathcal{P}_{2}(s)=\frac{\exp\!\!\left(\!\!-d_2\sqrt{\frac{s+k}{D}}\right)\!\!-\!\exp\!\!\left(\!\!-\!\left(2d_1+d_2\right)\!\sqrt{\frac{s+k}{D}}\right)}{s\!\!\left(\!\!1-\exp\left(-2(d_1+d_2)\sqrt{\frac{s+k}{D}}\right)\!\!\right)},
\end{align}
where $\mathcal{P}_{2}(s)$ is the Laplace transform of $P_{2}(t)$. To obtain the inverse Laplace transform of \eqref{laplace_transform}, we define two new equations as
\begin{align}
G(s)\!=\!\frac{\exp\!\left(\!\!-d_2\!\sqrt{\frac{s+k}{D}}\right)\!\!-\!\exp\!\left(\!-\left(2d_1+d_2\right)\!\sqrt{\frac{s+k}{D}}\right)}{\!(s+k)\!\left(\!1\!-\!\exp\left(-2(d_1+d_2)\!\sqrt{\frac{s+k}{D}}\right)\!\right)}
\end{align}
and
\begin{align}\label{H_s} H(s)\!=\!\!\underbrace{\frac{\exp\left(\!\!-\frac{d_2}{\sqrt{D}}s\!\right)}{s^2\!\!\left(\!\!1\!-\!\exp\!\left(\!-\frac{2(d_1+d_2)}{\sqrt{D}}s\right)\!\!\right)}}_{H_1(s)}\!-\!\underbrace{\frac{\exp\left(\!\!-\frac{2(d_1+d_2)}{\sqrt{D}}s\!\right)}{s^2\!\!\left(\!1\!-\!\exp\left(-\frac{2(d_1+d_2)}{\sqrt{D}}s\!\right)\!\!\right)}}_{H_2(s)}.
\end{align}

We note that $\mathcal{P}_{2}(s)=G(s)+\frac{k}{s}G(s)$ and $G(s)=H(\sqrt{s+k})$. Thus, we first solve the inverse Laplace transform of $H(s)$. From \eqref{H_s}, we observe that $H_1(s)$ and $H_2(s)$ have similar forms. Thus, we only show the process of performing the inverse Laplace transform of $H_1(s)$. We re-write $H_1(s)$ as
\begin{align}
	H_1(s)=&\exp\left(-\frac{d_2}{\sqrt{D}}s\right)\times\frac{1}{s\left(1-\exp\left(-\frac{d_1+d_2}{\sqrt{D}}s\right)\right)}\notag\\&\times\frac{1}{s\left(1+\exp\left(-\frac{d_1+d_2}{\sqrt{D}}s\right)\right)}.
\end{align}

According to \cite[eqs. (5.1), (5.34), (5.36), (1.18)]{oberhettinger2012tables}, the inverse Laplace transform of $H_1(s)$, denoted by $h_1(t)$, is 
 \begin{align}
h_1(t)&=\left\{\begin{array}{lr}
0, ~0<t<\frac{d_2}{\sqrt{D}}\\
(i+1)\left(t-\frac{d_2}{\sqrt{D}}-\frac{d_1+d_2}{\sqrt{D}}i\right), \\\frac{d_2}{\sqrt{D}}+\frac{2i(d_1+d_2)}{\sqrt{D}}<t<\frac{d_2}{\sqrt{D}}+2(i+1)\frac{d_1+d_2}{\sqrt{D}},
\\~~~~~~~~~~~~~~~~~~~~~~~~~~~~~~~~~~i=0,1,2,3,...
\end{array}
\right.
\end{align}

According to \cite[eq. (1.27)]{oberhettinger2012tables}, the inverse Laplace transform of $H_1(\sqrt{s})$ is
 \begin{align}\label{inverse_laplace_h_1}
 	&\mathcal{L}^{-1}\left\{H_1(\sqrt{s})\right\}=\frac{1}{2\sqrt{\pi t^3}}\int_{0}^{\infty}u\exp\left(-\frac{u^2}{4t}\right)h_1(u)\mathrm{d}u\notag\notag\\=&\frac{1}{\sqrt{4\pi t}}\sum_{i=0}^{\infty}\left(i+1\right)\left[2\exp\left(-\frac{u^2}{4t}\right)\left(\frac{id_1+(i+1)d_2}{\sqrt{D}}-u\right)\right.\notag\\&\left.+\sqrt{4\pi t}\mathrm{erf}\left(\frac{u}{\sqrt{4t}}\right)\right]\bigg|_{\frac{2id_1+(2i+1)d_2}{\sqrt{D}}}^{\frac{2(i+1)d_1+(2i+3)d_2}{\sqrt{D}}},
 	 \end{align}
where $F(x)\big|_a^b=F(b)-F(a)$.
As aforementioned, $H_1(s)$ and $H_2(s)$ have similar forms. Therefore, the inverse Laplace transform of $H_2(\sqrt{s})$, denoted as $\mathcal{L}^{-1}\left\{H_2(\sqrt{s})\right\}$, can be derived analogously. Based on \cite[eq. (1.3)]{oberhettinger2012tables}, \eqref{inverse_laplace_h_1}, and $\mathcal{L}^{-1}\left\{H_2(\sqrt{s})\right\}$, the inverse Laplace transform of $G(s)$, denoted by $g(t)$, is derived as
\begin{align}\label{G_s}
	g(t)=\exp\left(-kt\right)\left(\mathcal{L}^{-1}\!\left\{H_1(\sqrt{s})\right\}\!-\!\mathcal{L}^{-1}\!\left\{H_2(\sqrt{s})\right\}\right).
\end{align}

Given $\mathcal{P}_{2}(s)=G(s)+\frac{k}{s}G(s)$, the inverse Laplace transform of $\mathcal{P}_2(s)$ is
\begin{align}\label{final_equation}
	P_{2}(t)=g(t)+k\int_{0}^{t}g(u)\mathrm{d}u.
\end{align}

Substituting \eqref{G_s} into \eqref{final_equation}, we obtain \eqref{absorbed_molecules_at_receiver_2}. 

\section{Proof of Theorem \ref{theo2}}\label{app2}
According to the final value theorem, if $P_2(t)$ has a finite limit as $t\rightarrow\infty$, we have
\begin{align}\label{FVT}
\lim\limits_{t\rightarrow\infty}P_2(t)=\lim\limits_{s\rightarrow 0}s\mathcal{P}_2(s).
\end{align}

When $k\neq 0$, substituting \eqref{laplace_transform} into \eqref{FVT}, we obtain 
\begin{align}\label{neq}
P_{2,\mathrm{asy}}=\frac{\exp\left(-d_2\sqrt{\frac{k}{D}}\right)-\exp\left(-(2d_1+d_2)\sqrt{\frac{k}{D}}\right)}{1-\exp\left(-2(d_1+d_2)\sqrt{\frac{k}{D}}\right)}.
\end{align}

When $k=0$, we apply  L'H{\^o}pital's rule \cite{struik1963origin} to \eqref{neq}, and we have
\begin{align}\label{key}
P_{2,\mathrm{asy}}\!=&\!\lim\limits_{k\rightarrow 0}\!\!\frac{\frac{\partial}{\partial k}\left(\!\!\exp\left(-d_2\sqrt{\frac{k}{D}}\right)-\exp\left(-(2d_1+d_2)\sqrt{\frac{k}{D}}\right)\!\!\right)}{\frac{\partial}{\partial k}\left(1-\exp\left(-2(d_1+d_2)\sqrt{\frac{k}{D}}\right)\right)}\notag\\=&\frac{d_1}{d_1+d_2}.
\end{align}

Combining \eqref{neq} and \eqref{key}, we obtain \eqref{asy}.

\bibliographystyle{IEEEtran}
\bibliography{IEEEabrv,refs}
\end{document}